\documentstyle[twoside,fleqn,espcrc2,epsf]{article}

\newcommand{\AmS}{{\protect\the\textfont2
  A\kern-.1667em\lower.5ex\hbox{M}\kern-.125emS}}

\title{QCD Spectrum --- 1996}

\author{ 
Steven~Gottlieb\address{
Department of Physics, Indiana University, Bloomington, IN 47405, USA}%
\thanks{\mbox{Review talk at LATTICE96, St.~Louis, USA, June 1996}}
\hfill \raisebox{2.5cm}[0cm][0cm]{IUHET-342/August 1996}
}
       
\begin{document}
\thispagestyle{empty}

\begin{abstract}
Progress on the calculation of the spectrum from lattice calculations is
reviewed.  Particular emphasis is placed on discussing our ability to
control possible systematic errors coming from finite volume, and extrapolations
in quark mass and lattice spacing.  Recent approaches based on improved
actions are compared.

\end{abstract}

\maketitle

\section{INTRODUCTION}

A main goal of lattice QCD is to calculate the spectrum
of light hadrons\cite{RECENTREV}.  Such a calculation would not only be a major
achievement in its own right and a confirmation that QCD is the correct
theory of the strong interaction, it would give us added confidence in 
our ability to calculate many other nonperturbative quantities that are of
phenomenological interest.

It has been 15 years since the first pioneering calculations of the
spectrum were done \cite{PIONEERS} with computers capable of 
about 1 Megaflop, yet some
very simple questions are still relevant.  Can we control the systematic 
errors in lattice calculations?  Does the quenched or valence approximation
describe the real world?  Does QCD with dynamical quarks describe the real
world?  Can we improve the lattice action and ease the computational burden?
Are there hadrons in QCD that are not in the quark model, {\it e.g.}, glueballs
or exotics?  What are the quark masses?

Fortunately, the last question is the purview of a talk given earlier by
Paul Mackenzie \cite{MACKENZIE}.  
Subsequent sections of this paper discuss introductory
material regarding systematic errors, an overview of recent major simulations,
control of systematic errors, results for improvement schemes, and
calculations of glueballs and exotics.

\section{INTRODUCTION TO SYSTEMATIC ERRORS}

There are three {\it physical} sources of systematic errors in any
lattice calculation, the finite volume
$V$ or box size $N_s$, the quark mass $am_q$ which is always heavier
than in Nature, and the nonzero lattice spacing $a$.
In addition, there may be {\it algorithmic} sources of systematic error.
These would vary with the particular computational techniques used in
the calculation.  Chief among these are the use of the quenched or
valence approximation and the issue of whether Wilson or Kogut-Susskind
(aka, staggered or KS)
quarks are used.  {\it A priori}, we don't know how much difference it should
make to neglect the dynamics of the quarks.  In fact, this is one of our
basic questions.  On the other hand, the two quark representations have
different finite lattice spacing errors but are expected to have the same
continuum limit.  We would like to see this demonstrated by the calculations.
Additional possible sources of systematic error include convergence criterion
for iterative matrix inversions, gauge fixing accuracy when that is done,
and integration step size for molecular dynamics algorithms.  We will assume
these additional errors are all under control.

One important lesson of recent years is that very high statistics are needed
for careful study of systematic errors.  For the light quark spectrum,
the days of interest in qualitative (10\% error) calculations are long
past.  Researchers should be striving for mass values with errors of about
a fraction of a percent and errors in extrapolated quantities
of about 1--3\%.  This may seem like a high standard, especially
for more exploratory calculations with improved actions, but future progress
requires it.

\begin{figure}[thb]
\epsfxsize=0.99 \hsize
\epsffile{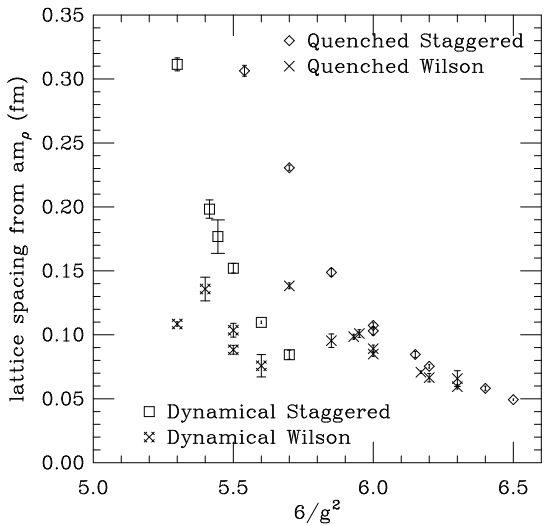}
\vspace{-28pt}
\caption{Lattice spacing as a function of $6/g^2$.}
\label{fig:spacing}
\end{figure}
\begin{figure}[th]
\epsfxsize=0.99 \hsize 
\epsffile{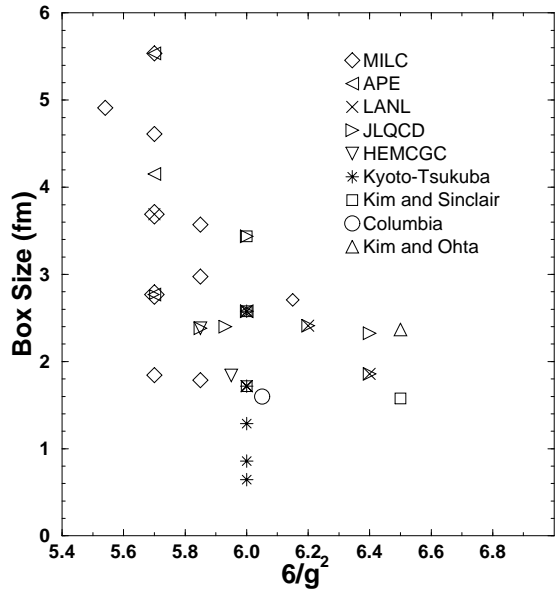} 
\vspace{-18pt}  
\caption{Box size {\it vs}. $6/g^2$ for quenched Kogut-Susskind calculations.}
\label{fig:quenched_ks}
\end{figure}

\section{OVERVIEW OF MAJOR SIMULATIONS} 
There have been a number of new calculations in the last year.  
To save space, we omit the traditional table of calculations, but direct
the reader to the database at the WWW site mentioned at the end of 
this paragraph.
We summarize the major simulations that
have been done (over approximately the last five years)
for the Wilson gauge action and either Wilson or
Kogut-Susskind quarks in a series of graphs.  In these graphs, we 
show the gauge coupling and the spatial size of the simulation.
The spatial size can be shown either in lattice units $N_s$ or in terms of
the physical size $a N_s$.
To save space here, we only show the physical size; however,
graphs of $N_s$ {\it vs}.\ $6/g^2$ as well as
many additional graphs are available on the WWW at {{\tt
{http://physics.indiana.edu/\~{}sg /lat96\_spectrum.html}}}.

To determine the physical size, we set the lattice
spacing by assuming that the $\rho$ mass at
zero quark mass is 770 MeV.  This involves an extrapolation in quark mass,
and the potential for introducing an error is discussed in the next
section.  In Fig.~\ref{fig:spacing}, 
we show the lattice spacing as a function of gauge
coupling.  We note that the large discrepancy between Wilson and
Kogut-Susskind scales decreases as the lattice spacing does.  Further, the
range of lattice spacing explored with dynamical quarks and in the quenched
approximation is by this measure not very different.
(However, this type of summary graph does not clearly
indicate the quality of the calculations.  For that, we need
to know more about the volume, quark masses studied and statistical quality of
the results.)

\begin{figure}[thb]
\epsfxsize=0.99 \hsize 
\epsffile{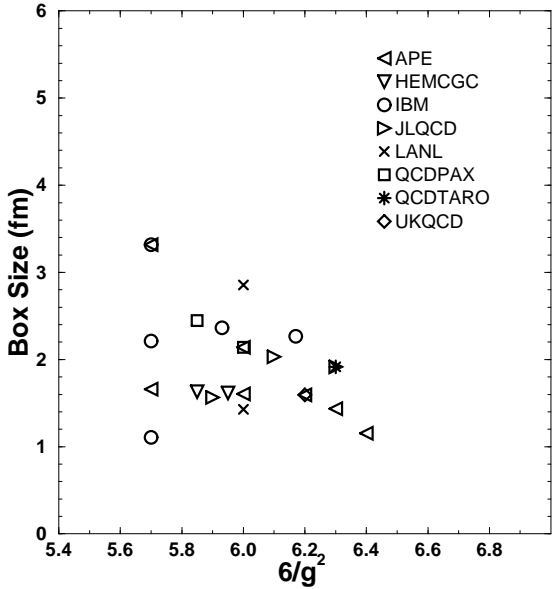} 
\vspace{-18pt}  
\caption{Box size {\it vs}. $6/g^2$ for quenched Wilson
calculations.}
\label{fig:quenched_wilson}
\end{figure}

\begin{figure}[thb]
\epsfxsize=0.99 \hsize 
\epsffile{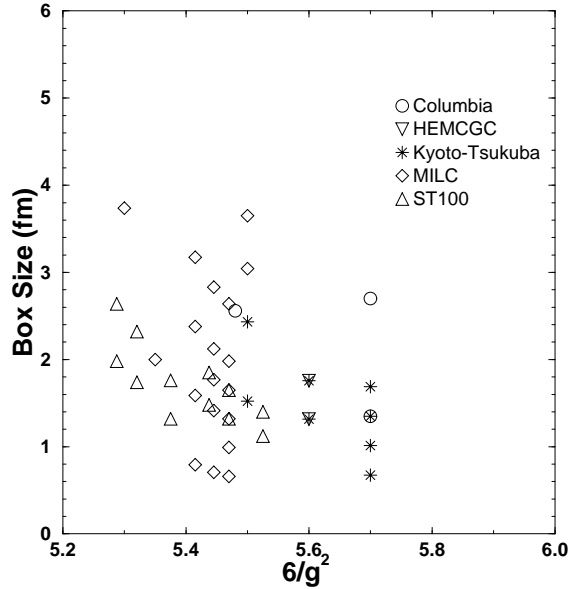}
\vspace{-18pt}  
\caption{Box size {\it vs}. $6/g^2$ for dynamical Kogut-Susskind
calculations.}
\label{fig:dynamical_ks}
\end{figure}

\begin{figure}[htb]
\epsfxsize=0.99 \hsize 
\epsffile{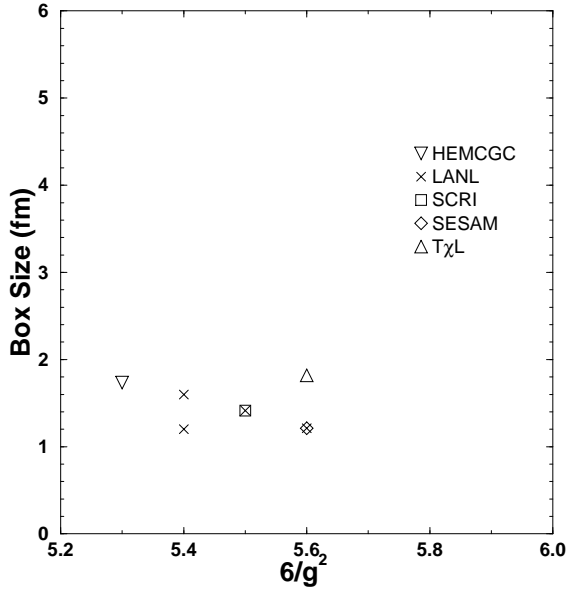}
\vspace{-18pt}  
\caption{Box size {\it vs}. $6/g^2$ for dynamical Wilson
calculations.}
\label{fig:dynamical_wilson}
\end{figure}

Figures~\ref{fig:quenched_ks}--\ref{fig:dynamical_wilson} 
show the box size as a function of $6/g^2$ for the major
simulations.  The WWW site contains color graphs where the color 
indicates the number of lattices analyzed in each calculation.
From these graphs one can immediately see that there is a general tendency
for the volumes to decrease as weaker couplings are used.  This leads us
into our discussion of systematic errors.


\section{CONTROL OF SYSTEMATIC ERRORS}
In this, the longest section of this paper we consider the three physical
sources of systematic error introduced in Sec.~2.  Where possible we try
to include results for quenched and dynamical, and Wilson and 
KS calculations.  In some cases, the results available from the literature,
including what was presented at Lattice '96, are insufficient to warrant
extrapolation to the physical limit.

\subsection{Finite size effects}
We expect that finite size (FS) effects will be quark mass
dependent, with the effect increasing as the quark mass decreases.  If we
had infinite computer resources, we would study hadron masses over a wide
range of volume, quark mass and coupling.  We could then see whether the
effect shows proper physical scaling.  That is, when we vary the coupling
do we find that the volume dependence is independent of $a$ for a fixed
physical quark mass (or $m_\pi/m_\rho$)?  With such detailed understanding,
we might actually be able to do accurate extrapolations in volume from
small volume calculations.  Unfortunately, we are far from this situation
and can only currently hope to identify a physical volume above which the
finite size effects are sufficiently small.

\begin{figure}[thb]
\epsfxsize=0.99 \hsize
\epsffile{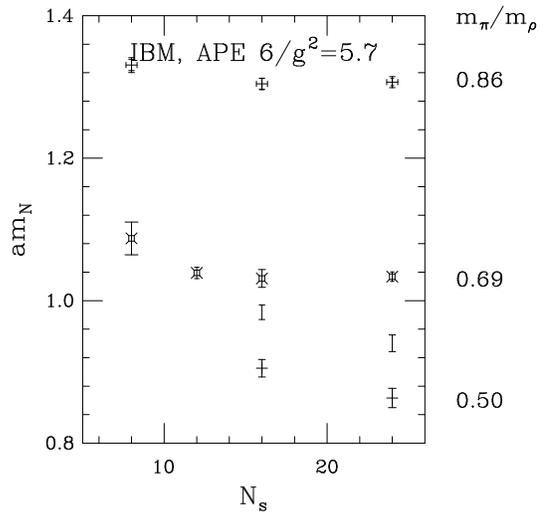}
\vspace{-28pt}
\caption{Nucleon mass as a function of spatial size for various quark masses.}
\label{fig:gf11nuc}
\end{figure}
\begin{figure}[htb]
\epsfxsize=0.99 \hsize
\epsffile{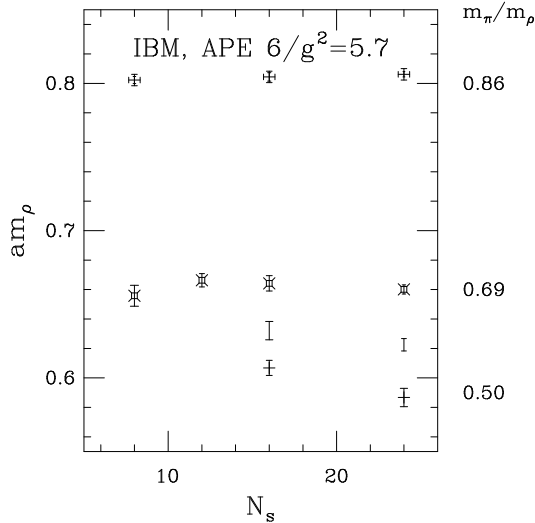}
\vspace{-28pt}
\caption{Rho mass as a function of spatial size.}
\label{fig:gf11rho}
\end{figure}

\begin{figure}[thb]
\epsfxsize=0.99 \hsize
\epsffile{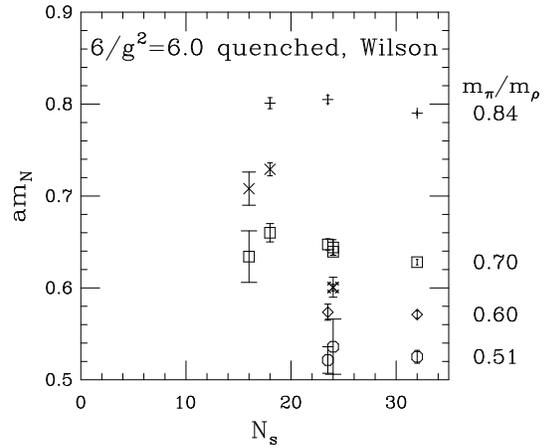}
\vspace{-28pt}
\caption{Nucleon mass as a function of spatial size for $6/g^2=6.0$.}
\label{fig:weakernuc}
\end{figure}

For quenched Wilson quarks, there have not been very extensive studies
of finite volume effects.  Only at $6/g^2=5.7$ and 6.0 are there any
calculations on more than two volumes.  Also, there tends to be more
variation in the hopping parameter $\kappa$
chosen by different groups than there is in
the staggered masses.  At $6/g^2=5.7$, the IBM group \cite{Butler94a}
has used three sizes
$N_s=8$, 16 and 24.  On those lattices, there are only two $\kappa$ values
that have
been used for all three volumes.  They correspond to $m_\pi/m_\rho=0.69$
and 0.86.  In Fig.~\ref{fig:gf11nuc}, we show their results for 
the nucleon mass along
with a result from the APE group for $N_s=12$ \cite{APE90}  
There does seem to be a FS
effect between $N_s=8$ and 16, for the heavier masses.  For the lightest
mass, $\kappa=0.1675$, the difference in the masses at 16 and 24 is
0.042(18).  This is a 2.6$\sigma$ or 4.8\% effect.  
(The next heavier
masses for $N_s=16$ and 24, are actually not at the same $\kappa$,
which is why we have no plotting symbol for those points.  
However, they are used in the chiral extrapolation, which was a 
linear extrapolation based on the three lightest 
quark masses \cite{Butler94a}.)
This is all the data we have for finite size effects for the nucleon at this
coupling.  
For the $\rho$, the data are shown in Fig.~\ref{fig:gf11rho}.  
There is no
evidence here for an effect at $N_s=8$.  For the lightest mass, there is
a difference of 0.028(8), which is a 2.5$\sigma$ or 3.3\% effect.
Using the $\rho$ to set the scale, the two largest lattices are 2.3 and 3.4
fm on a side.  Turning our attention to $6/g^2=6.0$, no single group has
done calculations on three volumes, but there are six calculations for
$N_s=16$
\cite{Daniel92a}, 18 \cite{APEnew}, 24 \cite{APE91c,QCDPAX96a,JLQCD95} 
and 32 \cite{LANL96}.  
The two largest
sizes here correspond to 2.1 and 2.9 fm.  For the nucleon, 
Fig.~\ref{fig:weakernuc} shows the data.  The mass
difference for the lightest $\kappa$ is 0.003(16).  Clearly, this is
consistent with no effect, but the error in the difference is 3\%, so we
have not ruled out the 3.3\% effect seen at 5.7.  At the next heaviest
$\kappa$, the difference is 0.0026(96) or a 1.6\% error in the difference.
At the heavier quark masses, there are some observable differences between
$N_s=24$ and 32.  For instance, for $\kappa=0.153$, the heaviest mass, the
difference is 0.0150(49) which is a 1.9\% or 3.1$\sigma$ effect.  At
$\kappa=0.155$, the difference is 0.0113(40) which is a 1.8\% or
2.8$\sigma$ effect.  
To save space, we do not include here the graph for the $\rho$.
We merely note that for the more precise calculation at $N_s=24$ the difference
between 24 and 32 is 0.010(6) which is a 2.5\% difference or 1.7$\sigma$.
If we only considered the $N_s=24$ result with the larger error bar, there
would be no observable effect.

\begin{figure}[thb]
\epsfxsize=0.99 \hsize
\epsffile{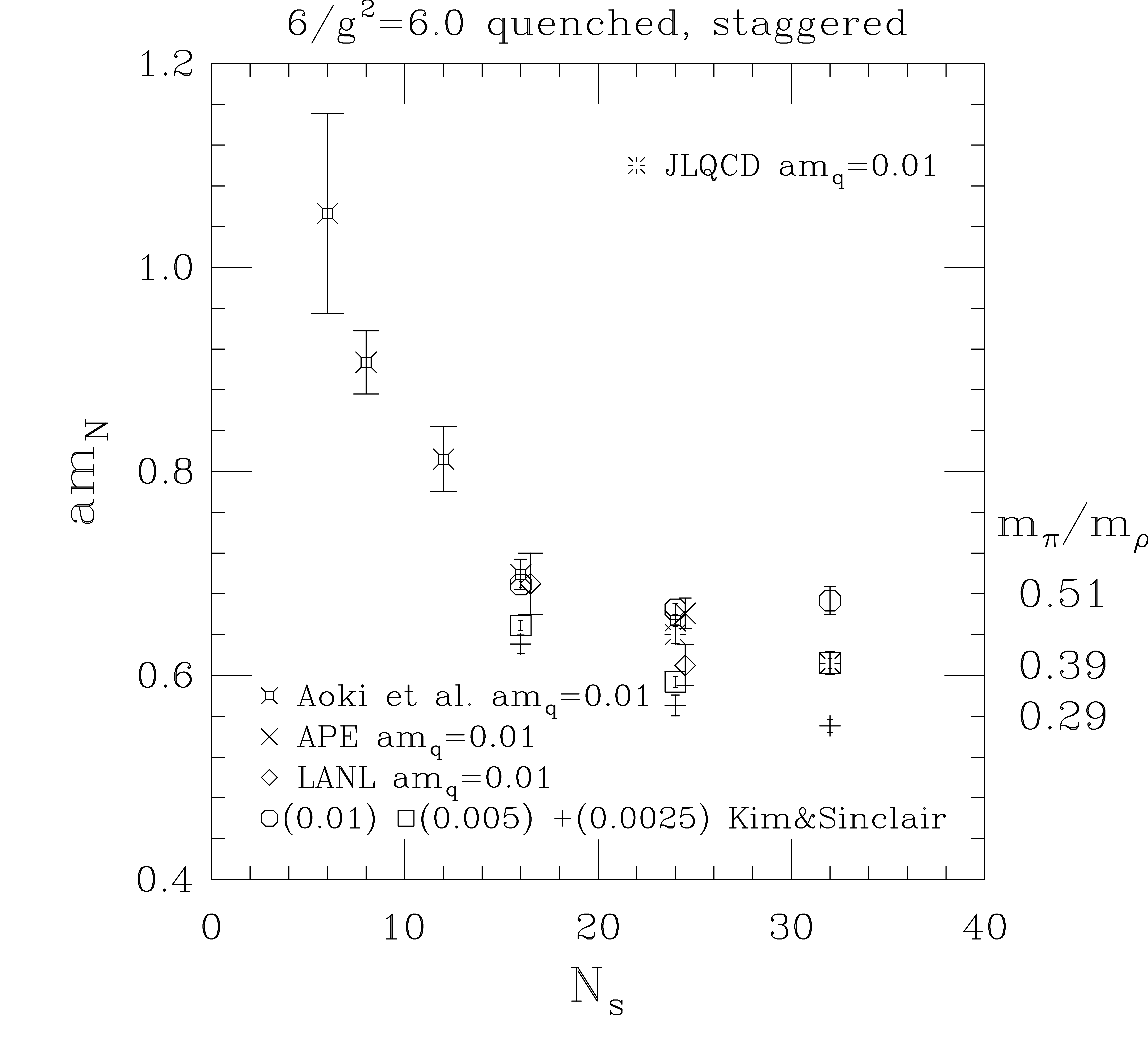}
\vspace{-28pt}
\caption{Kogut-Susskind nucleon mass as a function of spatial 
size for $6/g^2=6.0$.}
\label{fig:ks6.0nuc}
\end{figure}

For staggered quarks, there have been more extensive studies of finite size
effects in both the quenched \cite{Aoki94a,Gottlieb95a}
and dynamical cases\cite{Aoki94a,US}.
For $6/g^2=5.7$, six
lattice sizes have been studied from $N_s=8$ to 24.  For $6/g^2=6.0$,
six lattice sizes have been studied from 6 to 32.  In Fig.~\ref{fig:ks6.0nuc}, 
we show the
nucleon mass in lattice units {\it vs.}\ $N_s$.  The three largest sizes
correspond to 1.8, 2.6 and 3.5 fm.  The finite size effect is clearly quite
large at the smaller volumes, for which results are only available for the
heaviest quark mass $am_q=0.01$.  The heaviest quarks here ($m_\pi/m_\rho=0.51$)
are comparable to the lightest for the Wilson quark calculations.  Looking
at the octagons, we see that $am_N(N_s=16)-am_N(N_s=32)=0.0158(140)$ which is
slightly above 1$\sigma$, but represents a 2.2\% error in the difference.
Between 24 and 32, the difference is about 0.6$\sigma$ with the same size
error.  However, for the lightest quark mass,
$am_N(N_s=16)-am_N(N_s=32)=0.080(11)$ which is a 7$\sigma$ effect and one
standard deviation is also 2.2\% error.  Comparing only the two largest
sizes, the difference is 1.7$\sigma$ or 3.7$\pm$2.2\%.  It looks like we
have some good evidence for finite size effects with a 1.8 fm box,
especially at the lighter quark masses.  Between 2.6  and 3.5 fm, we don't 
have really strong evidence for an effect, but with an error in the
difference of about 2\%, there could be a few percent effect even on such
large volumes.

For dynamical staggered quarks, there has been extensive study of the
FS effects at two couplings\cite{Aoki94a,US,TSUKUBA}.  
This work is a few years old so we
merely recall that a box size of at least 2.5 fm was needed to eliminate
the FS effect for the quark masses studied there.  One
might need even bigger volumes for lighter quarks.  For dynamical Wilson
quarks, there are not enough results at different volumes to study this
issue.

To summarize, for the quenched approximation, we see FS
effects with a box size of under 2 fm.  For the lighter quark masses, there
might even be some effect on the nucleon mass for a box size as big as 2.5
fm.  A box size that large or greater is strongly recommended if the
nucleon mass is to be calculated.  Those who ignore this advice do so at their
own risk!
More high
statistics work is needed if we wish to determine what box size is needed
to reduce effects to the 1\% level for various quark masses.
We are far from understanding the effects well enough to
extrapolate from small volumes (say, 1.5 fm) to
the infinite volume limit with 1\% accuracy.

\subsection{Extrapolation in Mass}
The chiral extrapolation
may well now be the least well understood source of systematic error.
Chiral perturbation theory\cite{CHIRALPT} provides an expansion for the
hadron masses in terms of the quark mass; however,
in the quenched approximation\cite{CHIRALREV}, from the work of 
Bernard, and Golterman\cite{BERNARD}, and
Labrenz and Sharpe\cite{LABRENZ}, we 
know that there are additional terms, {\it e.g.}, a
$\sqrt m_q$ term for the nucleon, that are important at small quark mass.
The talk at this conference by Steve Sharpe\cite{SHARPE} 
deals in more detail with the difficulties of chiral extrapolation.

The chiral extrapolation process is complicated because chiral
perturbation theory is a small mass expansion, but our most accurate
numerical data are for large mass.  Thus, instead of extending our fits from
small mass to large mass and adding new chiral terms as needed, we are
often forced to use simple chiral forms (constant plus linear) for
relatively large mass and add additional terms when the simple forms don't
work.  

\begin{figure}[thb]
\vspace{-1cm}
\epsfxsize=0.99 \hsize
\epsffile{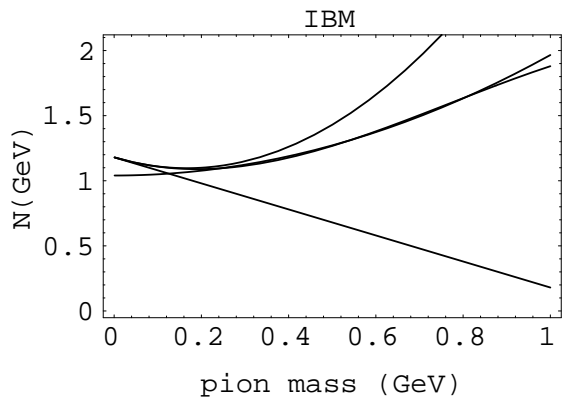}
\vspace{-28pt}
\vspace{-1cm}
\caption{Chiral extrapolation of the nucleon mass for
for $6/g^2=5.93$, including a two term fit to the intermediate mass region.}
\label{fig:nucleonfits}
\end{figure}

As an example of this problem, in Fig.~\ref{fig:nucleonfits}, we show several
curves related to the chiral behavior of the IBM group's nucleon mass at
$6/g^2=5.93$.  The curves come from a cubic fit by Gupta \cite{CHIRALREV}
and include the possibility
of a term proportional to $m_\pi$ that only appears in the quenched
approximation.  The three curves that converge at the $y$-axis are from
the Gupta fit to the data.  From top to bottom, they are the quadratic, cubic
and linear truncations of the fit.  
The fourth curve, which diverges from the others
at the $y$-axis is of the form $a+b m_\pi^2$, which is the lowest order 
contribution in chiral perturbation theory.  The coefficients $a$ and $b$
have been adjusted to fit the previous cubic fit over the 
range 0.45--0.85 GeV.
The two curves are nearly indistinguishable over a wider range, but at the
physical pion mass or below, there is clearly a significant difference.
We thus see that the extrapolation can be quite sensitive to which terms
are kept in the chiral approximation.

Over the past two years, several groups have been working
particularly hard to understand the chiral extrapolations.  
The LANL\cite{LANLREFS}, MILC\cite{MILCREFS}, and SCRI\cite{SLOANTALK}
collaborations have looked at vector mesons and the nucleon.  The
JLQCD collaboration\cite{JLQCDpion}, 
Kim and Sinclair\cite{KIM95}, Kim and Ohta\cite{KIMOHTA}, 
and Mawhinney\cite{MAWHINNEY} have
concentrated on the question of whether it is possible to observe quenched
chiral logarithms in the pion mass.

Among the first three groups mentioned above, the LANL and SCRI groups
have results for hadrons with unequal mass quarks; hence for the mesons
they have $n(n+1)/2$ mesons for $n$ quark masses.  Thus, they have more data
and more degrees of freedom when doing their fits.  This allows them more
freedom in choosing the range of quark mass included in the chiral fit.
MILC, on the other hand, has a very wide range of quark masses in its
calculations (a factor of 16 from lightest to heaviest). 
However, with five quark masses and only hadrons constructed from equal
mass quarks, there is limited freedom to play with the
range of mass included in the fits.  One of the difficulties in reviewing
the chiral extrapolations is that for quenched calculations 
hadron masses for different quark masses are correlated.  To get a proper
goodness of fit for the chiral extrapolation, those correlations must be
known, yet they are rarely published along with the hadron masses.
(It must be admitted, however, that not every group includes these
correlations in their chiral extrapolations.)

At Lattice '95, John Sloan \cite{SLOANTALK} presented evidence from the
SCRI collaboration that fits to the $\rho$ 
mass as a function of the $\pi$ mass can be extended to higher quark mass
if either either $C_3 m_\pi^3$ or $C_4 m_\pi^4$ is added to $C_0 + C_2
 m_\pi^2$.  For instance, with the cubic term, the range $0.58<m_\pi/m_\rho 
<0.93$ can be fit, but without it, the range is 0.58--0.77.  In terms of
$m_\pi$, the ranges are very roughly 0.5--1.3 GeV and 0.5--0.8 GeV, 
respectively.
Similar results have been seen in Refs.~\cite{LANL96,LANLREFS}, where the
$\rho$ was fit as a function of quark mass and a linear function was
sufficient to describe the data only over the range $0.5<m_\pi/m_\rho <0.7$.
With a higher power of quark mass, either $m_q^{3/2}$ or $m_q^2$, the fit
could be extended to $m_\pi/m_\rho=0.84$.  Neither group has been able to
clearly distinguish between the two higher powers; however, the first group
notes that a cubic seems to follow the data better than a quartic in the
high mass region where the fit is poor.

\begin{figure}[thb]
\epsfxsize=0.99 \hsize
\epsffile{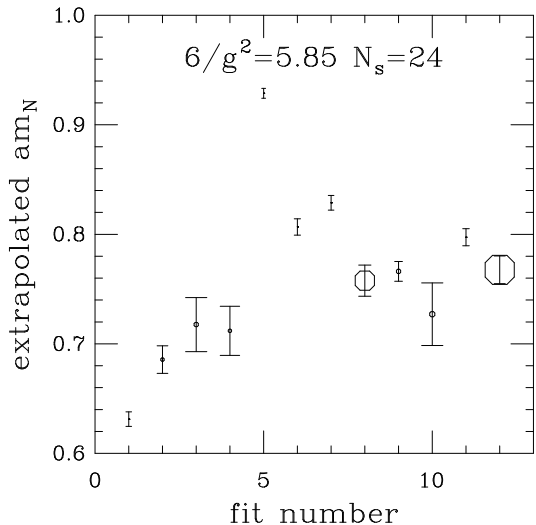}
\vspace{-28pt}
\caption{Kogut-Susskind nucleon mass extrapolated to the physical quark
mass for different chiral fits.  Size of plotting symbol is proportional to
the confidence level of the fit.}
\label{fig:milcfits}
\end{figure}

Turning to the Kogut-Susskind calculations, MILC\cite{MILCREFS}, with 
its wide range of
quark masses and small error bars, had great difficulty in fitting its
rho and nucleon masses with a single term in addition to $M_0 + b m_q$.
Twelve different fitting functions were studied.  They are up to four
parameter fits, and include terms such as $\sqrt{m_q} $ and $m_q \ln m_q$,
which are quenched chiral terms, as well as $m_q^{3/2}$, $m_q^2$ and 
$m_q^2 \ln m_q$, which are higher order terms that appear in both quenched
and ordinary chiral perturbation theory.  Figure~\ref{fig:milcfits} gives
an example of the variation of the nucleon mass with the choice of fitting
function.  The size of
the plotting symbols is proportional to the confidence level of
the fit.  Any visible symbol is a ``reasonable'' fit.  In the
presence of a $\sqrt{m_q} $ term, the nucleon mass decreases and the error
in the extrapolated value increases.
Looking at the other volumes and couplings, a combined confidence level
(CCL)
for all the cases can be calculated.  Adding to
$M_0 + b m_q$ (fit 5) a single power,
$m_q^{3/2}$ ($m_q^2$) gives a confidence of $3\times 10^{-10}$ ($10^{-24}$).
(These are fits 6 and 7).
There are six functions that have a CCL $>0.01$.
Five functions are four parameter fits, to fit the five masses.
The only three parameter fit, $M_0 + b m_q + c m_q \ln m_q$ (fit 9), has a
CCL of 1\%.
Among the four parameter fits, adding both $m_q^{3/2}$ and $m_q^2$ (fit 8) to
constant plus linear has the best CCL, 0.18, but is it not that much better 
than adding $\sqrt{m_q}$ in place of one of the higher powers, which give
0.12--0.13 (fits 2, 3).  
Fit 12 contains two higher order chiral terms, $m_q^2$ and $m_q^2 \ln m_q$.
Thus, it is not clear whether MILC may be seeing some evidence
for a quenched chiral effect in the nucleon mass.  

Quenched chiral logarithms have also been sought in the $\pi$ channel.
This particle has the advantage that its mass is the most precisely
determined.  The lowest order chiral prediction is
\begin{equation}
m_\pi^2 = A m_q .
\end{equation}
Kim and Sinclair\cite{KIMLAT93} did a quenched calculation
at $6/g^2=6.0$  and went to very light quark mass.
Kuramashi {\it et al}.\cite{KURAMASHI}
noticed the failure of the above relationship.
In subsequent work, Kim and Sinclair\cite{KIM95}
varied the lattice size to demonstrate control of
the finite size effects.  However, it was noted last year\cite{MILCREFS}
that at stronger
couplings than 6.0, the rise in $m_\pi^2/m_q$ as the quark mass
decreases occurs even for quite heavy quarks, thus making one wonder if the
rise is truly a chiral effect.  
Mawhinney\cite{MAWHINNEY} 
studied the $\pi$ mass and chiral condensate in both
quenched and dynamical quark configurations.  Varying the valence mass used
to make the measurements, he found linear behavior, but with a non-zero
intercept for $m_\pi^2$ and $\bar\chi\chi$.  This, he interpreted as a
FS effect, rather than a quenched chiral logarithm.

\begin{figure}[thb]
\epsfxsize=0.99 \hsize
\epsffile{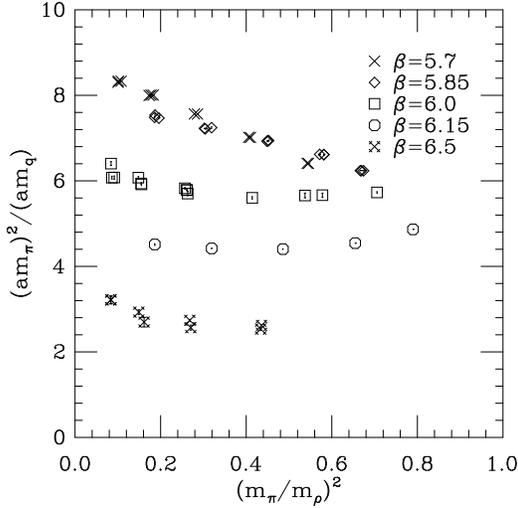}
\vspace{-28pt}
\caption{$m_\pi^2/m_q$ {\it vs.} $(m_\pi/m_\rho)^2$ for quenched staggered
calculations.}
\label{fig:mpisq_over_mq}
\end{figure}

Figure~\ref{fig:mpisq_over_mq} summarizes the
current results for couplings from 5.7--6.5.  The horizontal axis is
$(m_\pi/m_\rho)^2$ which makes it easier to compare different couplings
than in an earlier plot using the scale dependent $am_q$ on the $x$-axis.
Where there is data for different volumes is it all plotted to show the
FS effects.
It would certainly be valuable to have a lighter mass at 6.15 and heavier
masses at 6.5; however, it now looks like that at 6.0 and weaker coupling there
may be a broad flat region at intermediate quark mass.

\begin{figure}[thb]
\vspace{-0.6cm}
\epsfxsize=0.99 \hsize
\epsffile{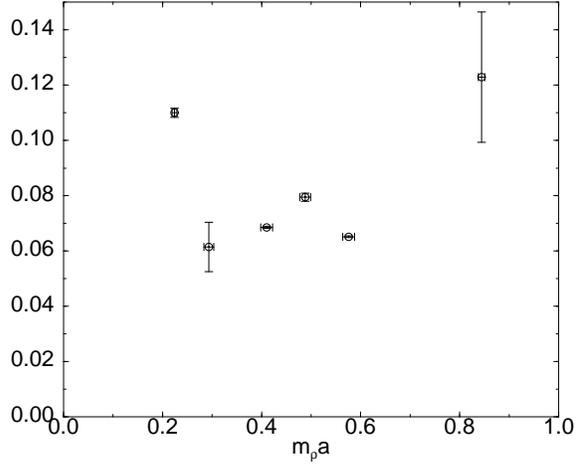}
\vspace{-28pt}
\caption{Coefficient $\delta$ of the quenched chiral logarithm {\it vs}.
$am_\rho$.}
\label{fig:jlqcddelta}
\end{figure}

\begin{figure}[thb]
\epsfxsize=0.99 \hsize
\epsffile{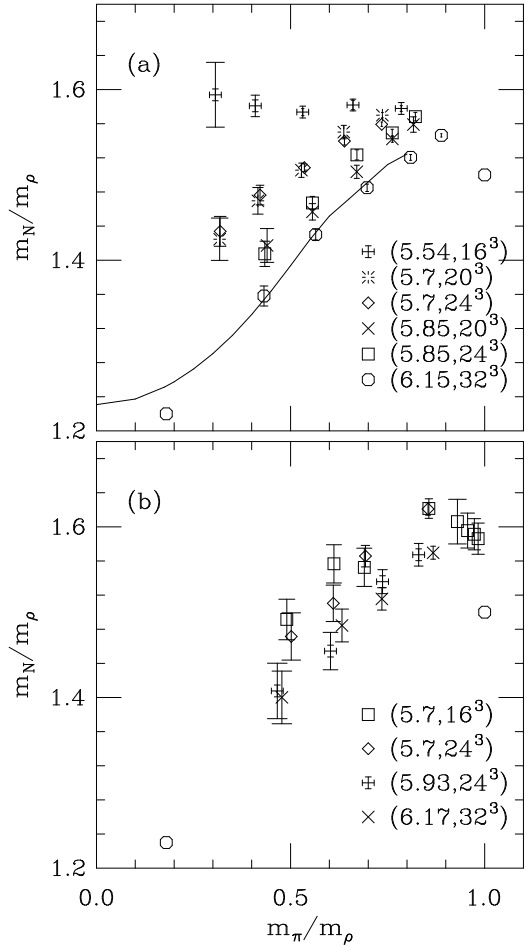}
\vspace{-38pt}
\caption{Edinburgh plot for quenched (a) KS quarks and (b) Wilson quarks.}
\label{fig:twoquedin}
\end{figure}
\begin{figure}[thb]
\epsfxsize=0.99 \hsize
\epsffile{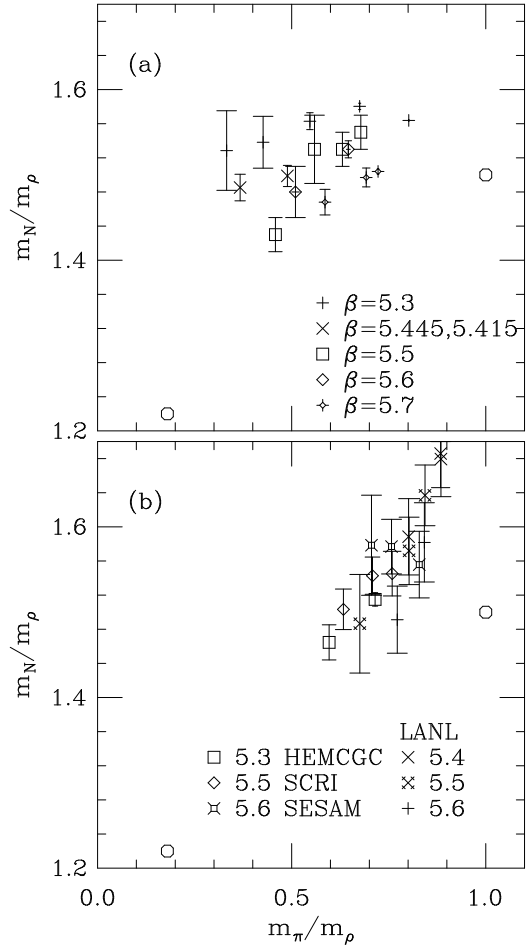}
\vspace{-38pt}
\caption{Edinburgh plot for dynamical (a) KS quarks and (b) Wilson quarks.}
\label{fig:twodynedin}
\end{figure}

The JLQCD collaboration\cite{JLQCDpion}
has presented a very interesting graph in which
they display the parameter $\delta$ of the chiral logarithm as a function
of lattice spacing.  They fit the $\pi$, including non-degenerate quark and
antiquark to the four parameter form 
\begin{eqnarray}
( m_\pi a )^2 &=& {(m_1+m_2)a \over 2}\cdot 2 A\cdot \nonumber \\
& & \!\!\!\!\!\!\! [1 - \delta \{ \log {2A m_1\over \Lambda}
+{m_2\over m_2-m_1}\log{m_2\over m_1}\} \nonumber\\
& & \mbox{} + C_4 (m_1+m_2)a ]
\end{eqnarray}
where
$A = {2 v a / f_\pi^2}$ and $\delta = {m_0^2}/(48\pi^2 f_\pi^2)$
with $v$ the chiral condensate.  In Fig.~\ref{fig:jlqcddelta}, we see that
$\delta$ is roughly 0.06--0.08 over the coupling range 5.85--6.2; however,
at 5.7 and 6.4, the computed value is higher.  Given that the computed
values seem to vary quite a bit, it may be too soon to say that the
quenched chiral logs have been observed in these calculations.

In summary, the chiral extrapolation, especially in view of possible
quenched chiral effects, remains an active area of investigation.  Many
simulations, particularly some of the older ones, contain only three or
four quark masses.  A wider range of quark masses, and getting hadron masses 
for non-degenerate quarks should help us to understand the chiral 
extrapolation better.  In comparing the results of different groups, it is
wise to make note of the range of quark mass or $m_\pi/m_\rho$ that is used
for the chiral fit.  Algorithmic improvements that enable us to more closely
approach the chiral limit would be valuable.  We must also caution that
if errors in the computed masses are overestimated, we may find good fits
to simple chiral forms in a region where higher order terms are necessary;
on the other hand, underestimation of errors may call for additional terms
to explain spurious variations that are merely statistical.

The Edinburgh plot remains a very valuable tool to summarize the results of
a spectrum calculation.  At the conference, eight such plots were shown
including three each for quenched staggered and Wilson calculations where
there is so much data that presenting it all on one graph would confuse
rather than enlighten.  The full set of graphs may be found at the WWW
site.  In Fig.~\ref{fig:twoquedin}, we present results from 
the MILC and IBM groups who have results at several couplings.  The upper plot
includes a curve showing the extrapolation to $a=0$.
In Fig.~\ref{fig:twodynedin}, we show results for dynamical quarks.
There are new results from SCRI\cite{SCRIDYN}, SESAM\cite{SESAM} 
and $T\chi L$\cite{TCHIL}
for dynamical Wilson quarks presented at this conference.

The quenched staggered simulations are generally going further toward the
chiral limit than the Wilson.  Dynamical Wilson calculations are very far from
the chiral limit ($m_\pi/m_\rho \ge 0.6$; staggered calculations have gone
closer to the chiral limit, but not at the weakest coupling studied (5.7).

\subsection{Extrapolation in lattice spacing}
\begin{figure}[thb]
\epsfxsize=0.99 \hsize
\epsffile{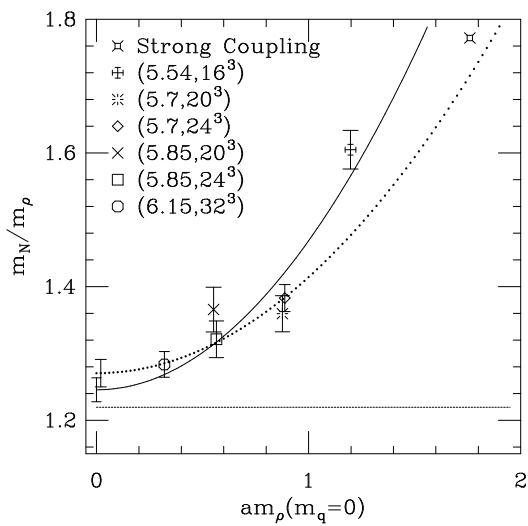}
\vspace{-28pt}
\caption{Extrapolation of $m_N/m_\rho$ as a function of $am_\rho$ for
quenched KS quarks.}
\label{fig:ksextrap}
\end{figure}

The extrapolation in lattice spacing is well understood compared to the
chiral case.  For Wilson quarks we expect errors of order $a$, while for
Kogut-Susskind quarks the errors should be of order $a^2$.  In
Fig.~\ref{fig:ksextrap}, we show the extrapolation to zero lattice spacing
based on data from the MILC collaboration.  The chiral extrapolations for
the rho and nucleon were based on the form $M+am_q+bm_q^{3/2}+cm_q^2$.
The strong coupling calculation at $6/g^2=5.54$ done this year and the
analytic strong coupling result of Kluberg-Stern {\it et al.}\cite{KLUBERG}
give confidence that the corrections are of order $a^2$.  
The two extrapolations either include or ignore the 5.54 data.  Only the
the largest volume data at 5.7 and 5.85 is used for the fit.
For dynamical
quarks, there are not sufficient results in the literature where both
FS effects and the chiral extrapolation are under control.  Thus, we make
no attempt to produce a similar plot.

\section{IMPROVEMENT} 

Quite a number of groups are attempting various improvement schemes.  
Improved or perfect actions\cite{LEPAGEREV,DEGRANDREV}
offer the hope of allowing lattice calculations on
much coarser lattices, which could save a tremendous amount of effort for
spectrum calculations, and possibly allow us to calculate new quantities
that are currently too costly to compute.
There were 59 contributions to the conference that had the string
``improve'' in the abstract.  Actually, one of them only claimed to have
improved statistics, but the others all dealt with some attempt to improve
the action beyond the usual Wilson gauge action and either Wilson or
Kogut-Susskind quarks.  Probably an entire hour would be insufficient
to review all of this work.   This is especially true in view of the wide
variety of calculations done.  I have tried to avoid having this
talk be a ``laundry list'' of each group's calculations.  When possible I
prepared summary graphs that show data from several groups.  With
the variety of improvements, it is hard to know which
calculations to compare.

Among the various improvement schemes, the one with the longest history is
the Sheikholeslami-Wohlert\cite{SW} or
``clover'' quark action with the Wilson gauge action.  Four groups have been
actively studying this scheme in the quenched
approximation.  
(Any calculation mentioned below without a reference should be assumed to be a
presentation at Lattice '96.)
Allton, Gimenez, Giusti and Rapuano\cite{APEnew} have
completed a series of runs at $6/g^2=6.0$, 6.2 and 6.4, with ensembles
ranging from 200 to 400 lattices.  
They also have
results for ordinary Wilson quarks.  Decay constants are an important focus
of this work.  Bhattacharya and Gupta have studied weak matrix elements
using the clover action, while Stephenson presented results with a
non-perturbative clover coefficient\cite{STEPHENSON}.
Finally, the UKQCD collaboration has
studied the couplings 5.7, 6.0 and 6.2, and has compared
tadpole-improvement with no improvement for 5.7.

The clover action in gauge configurations that include the effects of
dynamical Kogut-Susskind quarks has been studied by the SCRI
group consisting of Collins, Edwards, Heller and Sloan.  They use the old
HEMCGC lattices at $6/g^2=5.6$, $am_q=0.01$.
Borici and de Forcrand have investigated an improvement scheme based on
blocked lattices.

Recently, schemes based on improved gluonic actions have been extended to
include quarks.  Alford, Klassen and Lepage have investigated various
``highly improved'' quark actions.  The SCRI group mentioned above has
done an extensive series of calculations for six gauge couplings with the
clover action.  Fiebig and Woloshyn\cite{FIEBIG} 
have studied a quark action that
includes next-nearest neighbor interactions.  The MILC collaboration has
studied Kogut-Susskind quarks in improved glue, as well as a third nearest
neighbor interaction due to Naik\cite{NAIK}.  Finally, Morningstar and Peardon have
done glueball calculations using improved glue on an anisotropic lattice.
One of the potential difficulties with coarse lattices is that it may be
very difficult to fit the hadron propagators.  The plateau in an effective
mass plot is determined by physical considerations, so if it is 10 lattice
spacing on a lattice with $a=0.1$fm, it will only be 2.5 lattice spacings
if $a=0.4$fm.  With an anisotropic lattice that has a larger spatial
lattice spacing, it may be possible to reap most of the benefit in
computational cost without paying the price of not being able to fit the
masses.

Bock presented a nice talk in which he compared four different 
improvement approaches all at the coupling corresponding to the thermal
crossover for $N_t=2$.  He did spectrum calculations on $6^3\times16$
lattices and plotted ratios of $m_N$, $m_\Delta$ and $\sqrt{\sigma}$
to $m_\rho$, where $\sigma$ is the string tension.  
For the ``Cornell'' gauge action\cite{CORNELL}, 
he investigated three quark actions, Wilson,
clover and D234.  For the ``Iwasaki-Yoshi\'e'' gauge action\cite{IWASAKI} 
he used the D234 action.
He found good agreement for the ratios for all cases except the Wilson
quark action.

\begin{figure}[thb]
\epsfxsize=0.99 \hsize
\epsffile{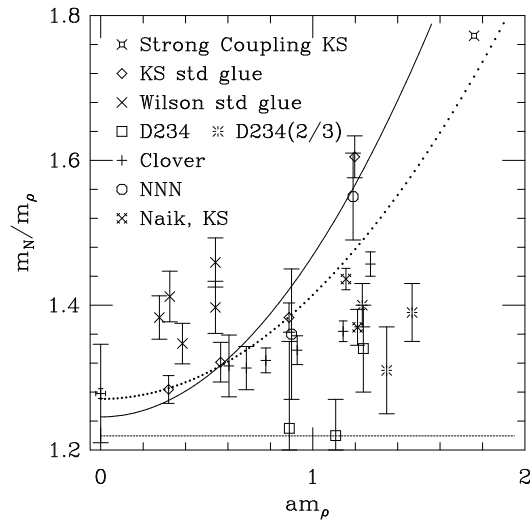}
\vspace{-28pt}
\caption{$m_N/m_\rho$ {\it vs}. $am_\rho$ for
quenched Wilson and KS quarks, and several improvement schemes.  Curves
come from previous graph for KS quarks.}
\label{fig:improvedextrap}
\end{figure}
Our final graph, Fig.~\ref{fig:improvedextrap},
compares results for $m_N/m_\rho$ from
a variety of quenched calculations.  For comparison, the results of 
Fig.~\ref{fig:ksextrap} are repeated.  The point above $am_\rho=0$ 
plotted with a fancy plus symbol is the
extrapolated value for Wilson quarks presented by the IBM group
(1.278$\pm$0.068). 
Their extrapolation was based on only three of the crosses.  From the left,
the first point is their result for $6/g^2=6.17$, the next is at 6.0 from
Bhattacharya {\it et al}., the next is IBM's result at 5.93.  The next two
points directly above each other are both 5.7 results from IBM.  They used
several sources and had two volumes at this coupling.  For the smaller
volume, the results are 1.459(34) (plotted) and 1.371(38), for combined
results of sink sizes 0, 1, 2 and for results from sink size 4 alone,
respectively.  On the larger volume, they found 1.397(36) (plotted) and
1.373(64), respectively.  The IBM continuum extrapolation was based upon
the smaller volume, sink 0, 1, 2 results.  (That volume is more comparable
to the physical volume of their weaker coupling calculations.)
Any of the other three results would clearly
decrease the fitted slope for $m_N/m_\rho$ and result in a larger
value of the ratio at $a=0$. Bhattacharya {\it et al}.\ have performed the
continuum extrapolation using their result and the sink 4, smaller volume
result and find 1.38(7) in the continuum limit.  Clearly, the 5.7 value
plays a crucial role in the extrapolation, and having more results at other
couplings would improve our understanding.

Results other than the KS points denoted by diamonds, Wilson points
denoted by crosses and the strong coupling KS result
all involve an improved gluon action.  The SCRI results, with six couplings,
comprise the most extensive results at this point.  The lattice spacing
dependence of the mass ratio is consistent with a quadratic correction.
It is not consistent with a linear dependence.  Results for the Wilson
quark action for these gauge configurations are also available.
Most of the other improved
glue results fall well below the KS result in ordinary glue (diamond above
$am_\rho=1.2$).  The results for the D234 and D234(2/3) actions from
Alford, Klassen and Lepage\cite{ALFORD} are promising, 
but the errors are too large to
determine the $a$ dependence.  The results from Fiebig and Woloshyn on the
next-nearest-neighbor action may indicate a stronger dependence on $a$, but
again it would be useful to have greater precision.

The two points denoted by a fancy cross compare the ordinary Kogut-Susskind
and Naik quark actions in the same improved glue configurations.
Their proximity indicates that most of the improvement comes from the glue,
not from the quark action.

\section{GLUEBALLS AND EXOTICS} 
There was not sufficient time in my talk for a discussion of glueballs or 
exotics; however, there were some interesting works that should be
advertized.  Bali presented a poster on new results for glueballs in
dynamical Wilson quark configurations.  Lee and Weingarten presented a talk
and poster session regarding scalar quarkonium and further evidence that
$f(1710)$ is a glueball.  Luo described recent glueball mass calculations
done in a Hamiltonian formalism.

The UKQCD collaboration has been studying hybrid mesons
recently\cite{UKHYBRID} and two papers
appeared shortly before the conference, one being a review by
Michael\cite{MICHAELREV}.  Toussaint presented a poster
session describing recent attempt of the MILC collaboration to calculate
exotic masses.

Also of iterest were two contributions related to the $\eta'$.  Massetti
described a bermion calculation of the $\pi$-$\eta'$ splitting and
Venkataraman described an $\eta'$ calculation done in configurations with
$N_f=0$, 2 and 4 dynamical staggered quarks.

\section{CONCLUDING REMARKS}
A great deal of interesting work is being done (much more than can be
adequately described here, so don't forget to check the WWW site).  We are on
the verge of answering some of the questions posed at the beginning of the
talk, especially as regards the quenched or valence approximation.  To
answer those questions will require additional work in order to demonstrate
control of systematic errors.  We need:
\begin{itemize}
\item Very high statistical accuracy ($<1$\%)
\item Large volumes
\item A wide range of quark masses, with special attention to the
chiral region and chiral extrapolations
\item Better understanding of relevant terms in the chiral
expansion.
\end{itemize}

Various improvement schemes are being pursued, but it is too soon to say
which approach is best.  There is certainly considerable evidence for a
faster approach to the $a\rightarrow0$ limit with several schemes.  
(Although it is not yet clear that all the schemes have the same limit!)
Improvement schemes are supposed to greatly ease the computational
burden.  If that is so, then a higher degree of statistical accuracy should
be expected in such studies.  Such accuracy as well as
reliable values for the computational effort
required for these calculations are necessary to decide which scheme is
best.

\vspace{5pt}
I gratefully acknowledge support by the U.S. Department of Energy under Grant
FG02-91ER 40661.  
I want to thank the organizers of Lattice '96 for inviting me to give
this review talk. 
I would also like to thank all the people who made
their results available to me before the conference.
I know that many of you were struggling to finish your analysis, just as I
was struggling to pull all the results together into a coherent picture.
I thank R.~Gupta and S.~Sharpe for discussions of chiral extrapolations, and
C.~Detar and
D.~Lichtenberg for reading the manuscript.  
Finally, I would like to thank all my MILC collaborators, especially
K.~Rummukainen who helped me prepare many of the color graphs.

\end{document}